\documentclass[preprint2]{aastex}
\usepackage{apjfonts}
\usepackage{epstopdf}
\usepackage{bm}
\def\Meszaros{M\'esz\'aros~}
\usepackage{epsfig}

\usepackage{url}

\begin{document}

\title{Origin of the GeV emission during
the X-ray flaring activity in GRB 100728A}

\author{Hao-Ning He\altaffilmark{1,2}, Bin-Bin Zhang\altaffilmark{3,4}, 
Xiang-Yu Wang\altaffilmark{1,2}, Zhuo Li\altaffilmark{5,6},  Peter \Meszaros\altaffilmark{3,7} }
\altaffiltext{1}{School of Astronomy and Space Science, Nanjing University,
Nanjing, 210093, China} \altaffiltext{2}{Key laboratory of Modern
Astronomy and Astrophysics (Nanjing University), Ministry of
Education, Nanjing 210093, China} \altaffiltext{3}{Department of
Astronomy and Astrophysics, Pennsylvania State University,
University Park, PA 16802, USA} \altaffiltext{4}{Department of
Physics, University of Nevada, Las Vegas, NV 89154, USA}
\altaffiltext{5}{Department of Astronomy, Peking University, Beijing
100871, China} \altaffiltext{6} {Kavli Institute for Astronomy and
Astrophysics, Peking University, Beijing 100871, China}
\altaffiltext{7}{Department of Physics, Pennsylvania State
University, University Park, PA 16802, USA}
\begin{abstract}
Recently,  Fermi-LAT detected GeV emission during the X-ray flaring
activity in GRB 100728A. We study various scenarios for its origin.
The hard spectrum of the GeV emission favors  the external
inverse-Compton origin  in which X-ray flare photons are
up-scattered by  relativistic electrons in the external forward
shock. This external inverse-Compton scenario, with anisotropic scattering effect
taken into account, can reproduce the temporal and spectral
properties of the GeV emission in GRB 100728A.

\end{abstract}


\section{Introduction}
One of the key discoveries by the {\it Swift} satellite is the
presence of X-ray flares during the early afterglow phase in a large
fraction of gamma-ray bursts (GRBs) (see, e.g., Burrows et al. 2005;
Zhang et al. 2006; Nousek et al. 2006). The rapid rise and decay
behavior of X-ray flares suggests that they are caused by internal
dissipation of energy due to late central-engine activity. It was
predicted that when the inner flare photons pass through the forward
shocks, they will be up-scattered by forward shock electrons,
producing a GeV flare (Wang et al. 2006; Galli \& Piro 2007; Fan et
al. 2008). The angular dispersion effect at the forward shock front
will wash out any shorter temporal structure (Beloborodov 2005), so
the high-energy inverse-Compton (IC) emission has a much smoother temporal structure
determined by the forward shock dynamical time. The incoming X-ray
flare photons are anisotropic as seen by the isotropically
distributed electrons in the forward shock, so the scatterings
between the flare photons and electrons in the forward shock are
anisotropic. This effect may decrease the IC emission in the
$1/\Gamma$ cone along the direction of the photon beam, where
$\Gamma$ is the bulk Lorentz factor of the forward shock, but
enhance the emission at larger angles, which leads to a delayed
arriving time of the IC photons relative to the flare photons (Wang
et al. 2006; Fan et al. 2008). On the other hand, the self
inverse-Compton emission of X-ray flares may give rise to a
high-energy component which should have similar light curves as the
flares (Wang et al. 2006; Fan et al. 2008; Yu \& Dai 2009). During
the early afterglows, the afterglow synchrotron emission can also
produce an extended high-energy component, which is thought to be
responsible for the decaying long-lived high-energy emission
detected from several GRBs by {\em Fermi}/LAT (e.g. Kumar \& Barniol
Duran 2009, 2010; Ghisellini et al. 2010; Wang et al. 2010; De
Pasquale et al. 2010; Corsi et al. 2010; He et al. 2011, Liu \& Wang
2011).

GRB 100728A is the second case (after GRB 090510) with simultaneous
detections by {\em Swift} and {\em Fermi}/LAT (Abdo et al. 2011). In
this paper, we first analyze the {\em Swift} XRT data and the {\em
Fermi}/LAT data of GRB 100728A, and obtain the X-ray and high-energy
($>$100 MeV) light curves (\S 2). The early (167-854 seconds) X-ray
afterglow exhibits intense and long-lasting flaring activity, with a
total of 8 flares (Abdo et al. 2011). Two X-ray flares are also seen
at the late stage (80-167 seconds) of the prompt phase (see Fig.1),
indicating that the X-ray activity continues from the prompt phase
to the afterglow phase. Interestingly {\em Fermi}/LAT detected
high-energy gamma-ray emission during the X-ray flaring activities
(Abdo et al. 2011). These simultaneous X-ray and GeV observations
offer a good case to study the nature of high energy emission. In
\S3, we confront the observations with the theoretical models and
find that the external inverse-Compton (EIC) scattering of X-ray
flare photons by electrons in the forward shock provides the best
explanation for the GeV emission. We calculate the light curve of
the EIC emission by taking into account the anisotropic scattering
effect, and compare it with that of the observed GeV emission by
{\em Fermi}/LAT in \S4. Finally, we give our conclusions in \S5.

\section{Observational facts of GRB100728A}

Bright GRB 100728A was triggered by both {\em Swift}/BAT and {\em
Fermi}/GBM with $T_{90}\sim 163$ s.
Intense X-ray flares are observed by {\em Swift}/XRT and significant
GeV photons are detected by {\em Fermi}/LAT in the meanwhile during
the early afterglow phase. Our study mainly focuses on this phase to
model both the X-ray and GeV behaviors. The XRT data are reprocessed
by our own codes (Zhang, B.-B et al. 2007). Pass 7 LAT data are
retrieved from the {\em Fermi} LAT data server and reprocessed using
a likelihood method. We selected "transient"(evclass=0) LAT photons
during the prompt emission phase and "source" (evclass=2) LAT
photons during the afterglow phase in a 20-degree circular region.
The time bins are judged according to the separation among the
prompt emission, flare and pure afterglow phases. Our results are
generally consistent with Abdo et al. (2011). The observational
facts related to our modeling in this work are summarized as
follows: i) The burst is bright with a fluence of $\sim1.3\times
10^{-4} {\rm erg cm^{-2}}$ in 10-1000 keV; ii) Ten successive flares
in XRT (from $\sim 80$ s to $\sim 854$ s) are detected, the
time-average spectrum of these flares can be described by a Band
function with $\alpha=-1.06\pm0.11$, $\beta=-2.24\pm0.02$ and a peak
energy $\varepsilon_{\rm pk}=1.0^{+0.8}_{-0.4}{\rm keV}$ (Abdo et
al. 2011); iii) The spectrum of the GeV emission is hard with a
photon index of $\Gamma_{\rm LAT}=-1.4\pm0.2$ (Abdo et al. 2011);
iv) The flux of GeV emission during the time $t\sim167-854$ s is
$F_{\rm LAT}\sim (5.8\pm4.5)\times10^{-9}{\rm
erg\,cm^{-2}\,s^{-1}}$. In what follows, we investigate which
theoretical scenarios can explain these observations.

\section{Confronting the observations with various models}

\subsection{Afterglow synchrotron emission scenario}
The underlying X-ray afterglow flux during the flare period should
be $F_{\rm X,af}\la 10^{-9} {\rm erg cm^{-2} s^{-1}}$ at time $t\sim
500$ s, according to Fig.1. The post-flare X-ray afterglow has
a decay index $\alpha_2=1.07\pm0.05$ before $t\sim10{\rm ks}$ and an
average photon index of $\Gamma_{\rm X}=-2.07\pm0.09$ (i.e.,
$\beta_2=-1.07\pm0.09$ for the spectral index in the convention
$F_{\nu}\propto t^{\alpha_2}\nu^{\beta_2}$) (Abdo et al. 2011, Evans
$\&$ Cannizzo 2010). Since the temporal index and spectral index
satisfy $\alpha_{2}\simeq\frac{3\beta_2+1}{2}$, i.e., the closure
relation for the afterglow emission in the fast-cooling
regime\footnote{Fast-cooling means that the electrons producing the
observed radiation cool down during the dynamic time. } for
constant-density interstellar medium (ISM) case (Zhang \& \Meszaros
2004). The power law index of electron number distribution $p$ (
i.e., $dn_{e}/d\gamma_{e}\propto\gamma_{e}^{-p}$), can  then be
obtained  $p=-2\beta_2\simeq2.2$
(Sari et al. 1998, Zhang $\&$ \Meszaros 2004). Thus, extrapolating
it to the LAT energies, the GeV flux from the forward shock should
be $F_{\rm GeV, af}=F_{\rm X,af}(\rm 1GeV/1keV)^{-\Gamma_{\rm
X}+2}\la0.6 \times10^{-9} {\rm erg\, cm^{-2}\, s^{-1}}$. This flux
is about one order of magnitude lower than the observed GeV flux,
disfavoring the forward shock synchrotron emission model. Moreover,
the hard spectrum of the GeV emission ($\Gamma_{\rm
LAT}=-1.4\pm0.2$, much harder than the X-ray spectrum with
$\Gamma_{\rm X}=-2.07\pm0.09$)  cannot be explained by this model.

\subsection{ X-ray flare self-IC scenario}
As argued in Abdo et al. (2011),  the LAT emission extends over the
flaring period in the early afterglow phase, rather than
mainly originated during the higher-significance flares. A
cross-correlation analysis between the LAT (diffuse-class) and XRT
light curves does not detect any significant temporal correlation or
anti-correlation between the two data sets (Abdo et al. 2011). These
facts, if true, would disfavor the self-IC scenario for the GeV
emission, since this scenario predicts a tight temporal correlation
between X-ray flares and GeV flares. However, we should note that
the GeV emission signal is not sufficiently strong to allow one to
draw a firm conclusion.

The hard GeV  spectral  shape can  be accounted  for by  a self-IC
component only if it peaks in or above the LAT energy window, i.e.
$\varepsilon_{\rm p, IC}\ga {\rm 1 GeV}$. As the peak energy of the
IC emission and the seed photon emission are related by
$\varepsilon_{\rm p, IC}\simeq2\gamma_{\rm m}^2\varepsilon_{\rm p,
syn}\simeq2\gamma_{\rm m}^2\varepsilon_{\rm pk}$ with the observed peak energy
$\varepsilon_{\rm pk}=1.0{\rm keV}$ (Abdo et al. 2011), this requires
$\gamma_{\rm m}\ga700$, where $\gamma_{\rm m}$ is the characteristic
Lorentz factor of the electrons.  Since the characteristic Lorentz
factor of electrons (in the comoving frame) is $\gamma_{\rm
m}=1.8\times10^3(p-2)/(p-1)\epsilon_e(\Gamma_{\rm sh}-1)$ (Wang et al. 2006),
where $\epsilon_e$ is the equipartition factor of electrons and
$\Gamma_{\rm sh}$ is the  Lorentz factor of the internal shock, this
would imply $\epsilon_e\ga0.4(p-1)/(p-2)(\Gamma_{\rm sh}-1)^{-1}$.
Since the relative shock Lorentz factor may be of order unity (Wang et al. 2006), i.e.
$\Gamma_{\rm sh}\sim 1-2$, the inferred value of $\epsilon_e$
is $\epsilon_e\ga1$ for $p\simeq 2-3$, which is too large to be
realistic. We note that this scenario can not be excluded if the
Lorentz factors ($\Gamma_{\rm sh}$) of the internal shocks for these
X-ray flares are very large.

\subsection{The external IC scenario}
The hard spectrum can be more easily accounted for in an external IC
(EIC) scenario, in which the  flare photons are scatterred by
external forward shock electrons. This is because the Lorentz
factors of the electrons in the forward shock are larger by a factor
of $\Gamma$ (i.e., the bulk Lorentz factor of the forward shock),
compared to the internal shock case,
and hence the EIC emission peaks at
much higher energies. In the adiabatic case, 
the shock Lorentz factor can be derived by
\begin{equation}\label{Gamma}
 \Gamma=44E_{53}^{1/8}n_0^{-1/8}t_{3}^{-3/8},
 \end{equation}
where we use the convention $Q_x=Q/10^{x}$ in cgs units hereafter
(Sari et al. 1998).
Adopting equation (\ref{Gamma}), we can obtain the typical Lorentz factor of the post-shock
electrons at time $t$ as $\gamma_{\rm
m}\simeq1.8\times10^3(p-2)/(p-1)\epsilon_e\Gamma=1.3\times10^3\epsilon_{e,-1}f_p
E_{53}^{1/8}n_{0}^{-1/8}t_{3}^{-3/8}$, where
$f_p\equiv\frac{6(p-2)}{(p-1)}$, $E$ is the kinetic energy of the
blast wave and $n$ is the number density of the circum-burst
medium.
By inputting the  value of the observed peak energy
$\varepsilon_{\rm pk}=1.0{\rm keV}$ (Abdo et al. 2011) into
$h\nu_{\rm m}^{\rm EIC}=2\gamma_{\rm m}^2\varepsilon_{\rm pk}$ (Sari
$\&$ Esin 2001, Wang et al. 2006), we can obtain the peak of the
observed EIC $\nu F_\nu$ flux as
\begin{equation}\label{numeic}
h\nu_{\rm m}^{\rm EIC}
=4\times10^{9}{\rm
eV}\epsilon_{e,-1}^{2}E_{53}^{1/4}n_{0}^{-1/4}t_{3}^{-3/4}
\end{equation}
for $p=2.2$.

The high flux of the X-ray flares will result in enhanced cooling of
the electrons in the forward shock. Adopting equation (\ref{Gamma})
and the shock radius as $R=4\Gamma^2 c t$(Waxman 1997), 
the energy density of  X-ray flare
photons in the forward-shock frame is
\begin{equation}\label{Ux}
\begin{array}{ll}
U_{\rm X}'=d_{\rm L}^2 F_{\rm
X}/(\Gamma^2 R^2c)\\
=0.34E_{53}^{-3/4}n_{0}^{3/4}F_{{\rm X},-8} d_{\rm
L,28}^{2}t_{3}^{1/4}{\rm erg\, s^{-1}\,cm^{-2}},
\end{array}
\end{equation}
where $F_{\rm X}$ is the observed flare flux, $d_{\rm L}$ is the
luminosity distance of the burst. Considering the case in which EIC
cooling is dominant, the cooling power is
$P(\gamma_e)=\frac{4}{3}\sigma_T c\Gamma^2\gamma_e^2U_{\rm X}'$
(Sari et al. 1998), then the cooling Lorentz factor of electrons
can be obtained by equation $\Gamma\gamma_c m_e c^2=P(\gamma_c)t$
(Sari et al. 1998), i.e.,
\begin{equation}\label{gammac}
\begin{array}{ll}
\gamma_{\rm c}\simeq3m_ec^2/(16
\sigma_{\rm T}cU_{\rm X}'\Gamma t)\\
=5\times10^2E_{53}^{5/8}n_0^{-5/8}F_{\rm X,-8}^{-1}d_{\rm
L,28}^{-2}t_3^{-7/8}.
\end{array}
\end{equation}
Thus, adopting equation (\ref{gammac}) and $\varepsilon_{\rm
pk}=1.0{\rm keV}$, the cooling break in the EIC spectrum can be
obtained as (Sari $\&$ Esin 2001, Wang et al. 2006)
\begin{equation}
\begin{array}{ll}
h\nu_{\rm c}^{\rm EIC}=2\gamma_{\rm c}^2\varepsilon_{\rm pk}\\
=5\times10^8{\rm
eV}E_{53}^{5/4}n_{0}^{-5/4}t_{3}^{-7/4}F_{\rm X,-8}^{-2}d_{\rm L,28}^{-4}.
\end{array}
\end{equation}

Below $\nu_{\rm m}^{\rm EIC}$, the EIC spectrum has a photon index
of $\Gamma_{\rm EIC}=-3/2$ if $\nu_{\rm c}^{\rm EIC}<\nu<\nu_{\rm
m}^{\rm EIC}$, or has the same index as the low-energy index
($\alpha=-1.06\pm0.11$) of the X-ray flare spectrum if
$\nu<\min(\nu_{\rm c}^{\rm EIC}, \nu_{\rm m}^{\rm EIC})$ (Sari $\&$
Esin 2001, Wang et al. 2006).
Therefore, below $\nu_{\rm m}^{\rm EIC}$, the photon index of the EIC emission
is consistent with the hard spectrum of the observed GeV
emission, $\Gamma_{\rm LAT}=-1.4\pm0.2$.

Using $R=4\Gamma^2ct$ for the radius of the forward shock and the
spectral peak $\varepsilon_{\rm pk}=1.0\rm keV$ (Abdo et al. 2011), we
can derive the peak spectral flux of the EIC emission as (Sari $\&$
Esin 2001)
\begin{equation}
\begin{array}{ll}
f_{\rm p}^{\rm EIC}=\tau_{\rm ssc}\left(\frac{F_{\rm
X}}{\varepsilon_{\rm pk}}\right) k_{a}=\frac{1}{3}\sigma_{\rm T}nR
\left(\frac{F_{\rm
X}}{\varepsilon_{\rm pk}}\right) k_{a} \\
=2\times10^{-33}k_{a}
E_{53}^{1/4}n_{0}^{3/4}t_{3}^{1/4}F_{\rm X,-8}{\rm
erg\,cm^{-2}\,s^{-1}\,Hz^{-1}},
\end{array}
\end{equation}
where $\tau_{\rm ssc}=\frac{1}{3}\sigma_{\rm T}nR$ is the
scattering optical depth of the forward shock shell, $\sigma_{\rm
T}$ is the Thompson cross section, $k_{a}$ is the correction factor
accounting for the suppression of the IC flux due to the anisotropic
scattering effect compared to the isotropic scattering case. It is
found that this correction is mild with $k_{a}\sim 0.4$ (Fan \&
Piran 2006; He et al. 2009).

Therefore, the peak flux of the EIC emission for the fast cooling
case (i.e., $\nu_{\rm c}^{\rm EIC}<\nu_{\rm m}^{\rm EIC})$ is (Sari
$\&$ Esin 2001)
\begin{equation}\label{numeicfmeic}
\begin{array}{ll}
\nu_{\rm m}^{\rm EIC} f_{\rm m}^{\rm EIC}=\nu_{\rm m}^{\rm EIC}f_{\rm p}^{\rm
EIC}\left(\frac{\nu_{\rm m}^{\rm EIC}}{\nu_{\rm c}^{\rm
EIC}}\right)^{-1/2} \\
=7\times10^{-10}k_a E_{53}\epsilon_{e,-1}t_3^{-1}d_{L,28}^{-2}{\rm
erg\,s^{-1}\,cm^{-2}}.
\end{array}
\end{equation}

From the spectrum of the observed GeV emission at $t=10^3{\rm s}$,
we have two constraints, i.e.,
\begin{equation}\label{numeiccon}
h\nu_{\rm m}^{\rm EIC}\ga 10^{9}{\rm eV},
\end{equation}
and
\begin{equation}\label{numeicfmeiccon}
\nu_{\rm m}^{\rm EIC} f_{\rm m}^{\rm EIC}\sim5\times10^{-10}{\rm
erg\,cm^{-2}\,s^{-1}}.
\end{equation}
Inputting equations (\ref{numeic}) and (\ref{numeicfmeic}) into the
above two constraints, respectively, one can get the following
constraints
\begin{equation}\label{emic}
n_0^{1/4}\la 4 E_{53}^{1/4}\epsilon_{e,-1}^{2} 
\end{equation}
and
\begin{equation}\label{eeE}
\epsilon_{e}\sim0.07k_a^{-1}E_{53}^{-1}d_{\rm L,28}^{2}. 
\end{equation}

The above constraints are consistent with the post-flare X-ray
afterglow observations.  We attribute the post-flare X-ray afterglow
to the synchrotron forward shock emission.  
Since its temporal index and spectral index satisfy the closure relation
$\alpha_2=\frac{3\beta_2+1}{2}$, as mentioned in section 3.1, 
the X-ray afterglow emission is in the fast cooling regime.
According to
equation (9) in Zhang et al. (2007), the synchrotron X-ray afterglow
flux is
\begin{equation}\label{nufxrt}
\begin{array}{ll}
\nu F_{\nu}^{\rm XRT}
=2\times 10^{-11}{\rm
erg\,s^{-1}\,cm^{-2}}
\\
\epsilon_{e,-1}^{p-1}\epsilon_{B,-2}^{\frac{(p-2)}{4}}E_{53}^{\frac{(p+2)}{4}}
\times g_{p}(1+Y)^{-1}t_4^{(2-3p)/4}\nu_{18}^{(2-p)/2}D_{\rm
L,28}^{-2},
\end{array}
\end{equation}
where $g_p=8.6\left(\frac{p-2}{p-1}\right)^{p-1}(3.3\times10^{-5})^{\frac{p-2.2}{2}}$, and the IC parameter is
$Y=(-1+\sqrt{1+4\eta\epsilon_e/\epsilon_B})/2$
with $\eta$ being the radiation efficiency (Sari $\&$ Esin 2001).
According to Sari et al. (1998),  the afterglow cooling Lorentz factor of electrons is 
$\gamma_c^{\rm af}=1.6\times10^3E_{53}^{-3/8}n_{0}^{-5/8}\epsilon_{B,-1}^{-1}t_4^{1/8}(1+Y)^{-1}$.
If the IC parameter $Y$ is not too large, we have $\gamma_c^{\rm af}>\gamma_m$ at $t\sim 10^4{\rm s}$, 
implying a slow cooling case, where the radiation efficiency is  
$\eta=(\gamma_c^{\rm af}/\gamma_m)^{2-p}=0.8h_pE_{53}^{2-p}n_0^{(p-2)/2}\epsilon_{B,-1}^{p-2}\epsilon_{e,-1}^{p-2}t_4^{(2-p)/2}(1+Y)^{p-2}$
with $h_p=1.1^{2.2-p}$ (Sari $\&$ Esin 2001).
At $t=10^4$ s, for $D_L=2\times10^{28}{\rm cm}$ (i.e., at
redshift $z$=1), with $\epsilon_e\sim 0.3$, $\epsilon_B\sim0.1$, $n=1{\rm cm^{-3}}$ and $E\sim3\times10^{53} {\rm
erg}$, the IC parameter is $Y\simeq 1$.
For the above parameters,  the flux derived from equation (\ref{nufxrt}) is consistent
with the observed X-ray flux $\nu F_{\nu}^{\rm
XRT}\sim3\times10^{-11}{\rm erg\,cm^{-2}\,s^{-1}}$.

Note that in the above estimate we have implicitly assumed that the
X-ray flare flux is sufficiently strong that the electrons in the
forward shocks are cooled down by the flare photons (i.e.
$\gamma_{\rm m}>\gamma_{\rm c}$). Nevertheless, in the regime that
electrons cool slowly ( i.e., $\gamma_{\rm m}<\gamma_{\rm c}$ in the
slow cooling case), one can also have an EIC spectrum as hard as the
low-energy spectrum of the X-ray flare (i.e., $\Gamma_{\rm
EIC}\simeq\alpha=-1.06\pm0.11$) if $h\nu_{\rm m}^{\rm EIC}\ga 1 {\rm
GeV}$. Such a situation  will be included in the numerical modelings
in \S 4.

\subsection{The afterglow synchrotron self-Compton emission scenario }
Predictions were made that the synchrotron self-Compton (SSC)
emission from the forward shock electrons can produce  GeV emission
during the early afterglow in some parameter space (e.g. Zhang \&
\Meszaros 2001). In the case of GRB 100728A, however, the
illumination by X-ray flare photons enhances the cooling of the
forward-shock electrons, which in turn suppresses the afterglow
synchrotron and SSC emission. We expect that the ratio of the EIC
luminosity  to the SSC luminosity  is ${L_{\rm EIC}}/{L_{\rm
SSC}}={U'_{\rm X}}/{U'_{\rm syn}}$, where $U'_{\rm X}$ and ${U'_{\rm
syn}}$ are, respectively, the energy density of flare photons and
synchrotron photons in the comoving frame of the forward shocks.
During the flare activity period, because the flux of flare photons
is significantly larger than that of the synchrotron X-ray
afterglow, i.e. ${U'_{\rm X}}>{U'_{\rm syn}}$, the SSC emission
should be subdominant compared to the EIC emission.

\section{Numerical modeling of the GeV emission}
In this section we calculate numerically the flux of the EIC emission
of the x-ray flare photons scattered by forward shock electrons, taking
into account the anisotropic scattering effect. For a photon beam
penetrating into the shock region where electrons are isotropically
distributed,
the EIC emissivity of
radiation scattered at an angle $\theta_{\rm SC}$ relative to the
direction of the photon beam in the shock comoving frame
is (Aharonian $\&$ Atoyan1981, Brunetti 2000, Fan et. al. 2008,
He et al. 2009):
\begin{equation}\label{AIC}
\begin{array}{ll}
\varepsilon'^{\rm EIC}(\nu',\cos\theta_{\rm
SC})\approx\frac{3\sigma_{\rm T}c}{16\pi}
\int^{\gamma_{\rm max}}_{\gamma_{\rm m}}d\gamma_e
\frac{dn_e}{d\gamma_e}\int^{\nu'_{\rm s,max}}_{\nu'_{\rm s,min}}
\frac{f'^{\rm X}_{\nu'_{\rm s}}d\nu'_{\rm s}}{\gamma_e^2\nu'_{\rm s}}\\
\left[1+\frac{\xi^{2}}{2(1-\xi)}-\frac{2\xi}{b_{\theta}(1-\xi)}+\frac{2\xi^2}{b_{\theta}^2(1-\xi)^2}\right],
\end{array}
\end{equation}
where $\nu'_{\rm s}$ and $\nu'$ are respectively the comoving-frame
frequencies of the photons before scattering and after
scattering,
$\gamma_e$ is the Lorentz factor of scattering electrons,
$\xi\equiv h\nu'/(\gamma_em_ec^2)$, $b_{\theta}\equiv
2(1-\cos\theta_{\rm SC})\gamma_eh\nu'_{\rm s}/(m_ec^2)$,
and $h\nu'_{\rm s}\ll h\nu'\ll \gamma_e
m_ec^2b_{\theta}/(1+b_{\theta})$.
Integrating equation (\ref{AIC}) over the angle $\theta_{\rm SC}$
for the whole solid angle, one can reduce equation (\ref{AIC}) to
the equation for the case of isotropically distributed electrons and
photons. The maximum Lorentz factor of electrons $\gamma_{\rm
max}$ 
is obtained by
equating the electron acceleration timescale to the cooling
(including synchrotron cooling and IC cooling) timescale.
$f_{\nu'_{\rm s}}^{'\rm X}$ is the flux density of seed photons
at the frequency $\nu'_{\rm s}$, which are X-ray flare photons in
our case. The lowest and highest frequencies of seed photons in the
shock frame can be calculated via $\nu'_{\rm s,min}=0.3{\rm
keV}/\Gamma$ and $\nu'_{\rm s,max}=10{\rm keV}/\Gamma$, according to
the observation range of Swift XRT, i.e., $0.3{\rm keV}-10{\rm
keV}$. The light curves of these X-ray flares are modeled by
power-law rises and decays in the calculation, as shown by the green
lines in Fig.1 and Fig.2. The spectra of these X-ray flares are
modeled by the Band function as given in section 2.

The  electron distribution $dn_e/d\gamma_e$ in Eq.(13) depends on
where the cooling break $\gamma_{\rm c}$ is. When the circum-burst
density is sufficiently low,  the forward shock electrons suffer
from weak cooling by X-ray flares,  so we have $\gamma_{\rm
m}<\gamma_{\rm c}$. Otherwise, we have $\gamma_{\rm m}>\gamma_{\rm
c}$. The distribution of electrons is given by (Sari et al. 1998)
\begin{equation}
\frac{dn_{e}}{d\gamma_{e}}\propto\left \{\begin{array}{ll}
\gamma_{e}^{-2},&\gamma_{\rm c}\leq\gamma_{e}\leq\gamma_{\rm m}\\
\gamma_{e}^{-p-1},&\gamma_{\rm m}<\gamma_{e}<\gamma_{\rm max}\end{array}\right.
\end{equation}
 for $\gamma_{\rm c}\leq\gamma_{\rm m}\la\gamma_{\rm max}$, and
\begin{equation}
\frac{dn_{e}}{d\gamma_{e}}\propto\left\{\begin{array}{ll}
\gamma_{e}^{-p},&\gamma_{\rm m}\leq\gamma_{e}\leq\gamma_{\rm c}\\
\gamma_{e}^{-p-1},&\gamma_{\rm c}<\gamma_{e}<\gamma_{\rm max}\end{array}\right.
\end{equation}
for $\gamma_{\rm m}<\gamma_{\rm c}\la\gamma_{\rm max}$.

The flux  density  of the  EIC emission at a frequency $\nu$ is given by
(Huang et al. 2000, Yu et al. 2007)
\begin{equation}
F_{\nu}^{\rm EIC} =\int^{\theta_{\rm j}}_{0}\frac{\varepsilon'^{\rm
EIC}(\nu/D,\cos\theta)D^2}{d_{\rm L}^2} 2\pi
R^3\sin{\theta}\cos{\theta}d\theta,
\end{equation}
where $D=1/[\Gamma(1-\beta \cos\theta)]$ is the Doppler factor,
$\beta$ is the velocity of the forward shock, $\theta$ is the angle
between the motion of the emitting material and the jet axis, and
$\theta_{\rm j}$ is the opening angle of the jet. Since $\theta$
also represents the angle between the injecting photons and the
observed scattered photons in the observer frame, it is related with
the angle $\theta_{\rm SC}$ in the comoving frame by
$\cos\theta_{\rm SC}=(\cos\theta-\beta)/(1-\beta\cos\theta)$.
$\theta_{\rm j}$ is a function of time with an initial value of
$\theta_{\rm j,0}$. Since the integration of Eq. (13) is performed
over the equal arrival time surface (EATS), the radius $R$ of the
shock is a function of $\theta$, which is determined by
\begin{equation}
t=\int\frac{1-\beta \cos{\theta}}{\beta c}dR\equiv {\rm constant}
\end{equation}
within the jet boundaries (e.g. Waxman 1997; Granot et al. 1999).

The calculated light curves and spectra of the EIC emission are
shown in Fig.1 and Fig.2 for high ($n=1 {\rm cm^{-3}}$) and low
($n=0.01 {\rm cm^{-3}}$) circumburst density cases respectively, in
which the electron distribution is, correspondingly, in the
fast-cooling and slow-cooling regime. In both cases, the EIC
emission is the dominant component at GeV energies during the X-ray
flare activity period. In the high density case (Fig.1), the
electrons are in the fast cooling regime during the X-ray flare
period and we have $\nu_{\rm c}^{\rm EIC}<\nu_{\rm LAT}<\nu_{\rm
m}^{\rm EIC}$, leading to a hard GeV emission spectrum with a photon
index $\Gamma_{\rm GeV}=-1.5$, as  shown by the inset spectra of
Fig. 1. In contrast to the high temporal variability of the x-ray
flares, the EIC emission is very smooth. It is clearly seen that the
EIC emission starts later than the earliest flares, which can
explain the non-detection of GeV emission during the earliest two
flares (80-167s). The EIC emission continues after the X-ray flare
activity shuts off, which  may explain the marginal detection of GeV
emission after the X-ray flare activity, as reported in Abdo et al.
(2011).

Fig.2 shows the case of a low circumburst density with $n=0.01 {\rm
cm^{-3}}$. In this case, a large shock radius results in a lower
flare photon density in the shock comoving frame so that the EIC
cooling is not important. In this case, we will have
$\gamma_m<\gamma_c$, and thus $\nu_{\rm LAT}<\nu_{\rm m}^{\rm
EIC}<\nu_{\rm c}^{\rm EIC}$. The spectrum of the EIC emission below
$\nu_{\rm m}^{\rm EIC}$ has the same slope as that of the seed
photons, i.e. the photon index is similar to the low-energy photon
index $\alpha$ of the Band function describing the spectrum of X-ray
flares. Such a spectrum is also consistent with the measured
spectrum of the GeV emission within the error bars, as shown in the
inset plot of Fig.2. In the low density case, the afterglow
synchrotron emission dominates over the SSC emission in the GeV band
in the early time.

\section{Conclusions}
For the first time, GeV emission from a GRB during the X-ray flaring
activity has been detected by the {\em Fermi}/LAT, in GRB100728A.
The temporal coincidence between the GeV emission and the X-ray
flares suggests that the GeV emission should be related to the
flares in some way. Here we have shown that an EIC scenario, where
X-ray flare photons are up-scattered by electrons  in the external
forward shocks, provides the best explanation for the GeV emission,
supporting the earlier prediction of GeV emission from GRB X-ray
flares (Wang et al. 2006). The hard spectrum of the GeV emission can
be readily explained by the EIC emission. The delayed behavior of
the GeV emission relative to the X-ray flares  also naturally arises
in the EIC scenario. Synergistic observations between {\em Swift}
and {\em Fermi} should be able in the future to find other similar
cases, allowing further tests of this EIC scenario.

We are grateful to Bing Zhang and Zigao Dai for valuable
discussions. This work is supported by the NSFC under grants
10973008 and 11033002, the 973 program under grant 2009CB824800, the
program of NCET, the foundation for the Authors of National
Excellent Doctoral Dissertations of China, Jiangsu Province Innovation for PhD candidate CXZZ11$\_$0031, the Fok Ying Tung
Education Foundation, NASA NNX08AL40G and NSF PHY 0757155. It is
also supported in part by NASA via the Smithsonian Institution grant
SAO SV4-74018,A035 [BBZ] at PSU.

\clearpage
\begin{figure*}
\plotone{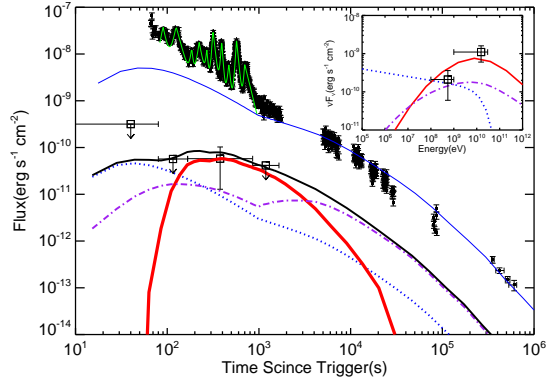} \caption{Light curves of the GeV emission and
X-ray afterglow of GRB100728A. The filled black squares represent
the observed X-ray flux by Swift XRT (0.3-10{\rm keV}) and the empty
black squares represent the high energy data and upper limits from
Fermi/LAT ($100 {\rm MeV}-30{\rm GeV}$). The green solid line
represents the input light curve of X-ray flares that we used in the
calculation. The blue solid line represents the  synchrotron
afterglow emission in 0.3-10 keV, while the red solid line
represents the high energy emission (in $100 {\rm MeV}-30{\rm GeV}$)
produced by the external inverse-Compton scattering of the X-ray
flare photons. The blue dotted line and purple dash-dotted line
represent, respectively, the afterglow synchrotron emission and the
afterglow SSC emission in $100 {\rm MeV}-30{\rm GeV}$. The black
solid line is the sum of the three high energy components
(represented by the red solid line, blue dotted line and purple
dash-dotted line, respectively). For clearness, we re-scaled the
flux of all the high-energy components by multiplying a factor of
$0.01$. The inset plot shows the spectra of the three high-energy
components  and the LAT data taken from Abdo et al. (2011) during
the X-ray flaring period (167-854 s). $E=3\times10^{53}{\rm erg},
n=1{\rm cm^{-3}}, p=2.2, \Gamma=250, \theta_{\rm j,0}=0.1,
\epsilon_{e}=0.1, \epsilon_{B}=0.04$ and $z=1$ are used in the
calculation.}
\end{figure*}

\begin{figure*}
\plotone{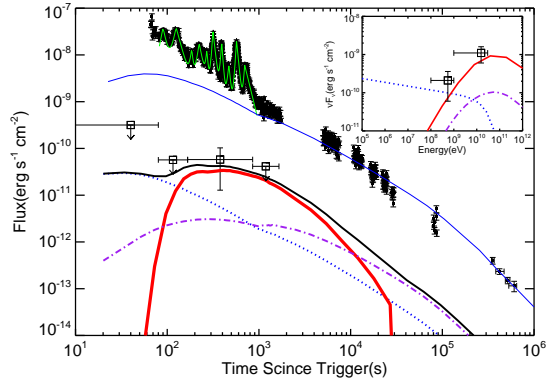}\caption{The same as Figure 1, but for a
low-density case ($n=0.01{\rm cm^{-3}}$).  $E=1\times10^{53}{\rm
erg}, n=0.01{\rm cm^{-3}}, p=2.2, \Gamma=400, \theta_{j,0}=0.07,
\epsilon_{e}=0.2, \epsilon_{B}=0.1$ and $z=1$ are used in the
calculation. }
\end{figure*}

\end{document}